\documentclass[twocolumn,pra,aps,amsmath,amssymb,amsfonts,showpacs]{revtex4}
\usepackage{graphicx}
\usepackage{bm}
\usepackage{color}
\newcommand{\tst}{\textstyle}

\newcommand{\mbf}{\mathbf}
\newcommand{\mrm}{\mathrm}

\newcommand{\squiz}[1]{\!#1\!}

\begin{document}

\title{Shape resonances in modified effective-range theory for electron-molecule collisions}
\author{Zbigniew Idziaszek}
\affiliation{Instytut Fizyki Teoretycznej, Uniwersytet Warszawski, 00-681 Warszawa, Poland}
\affiliation{Centrum Fizyki Teoretycznej, Polska Akademia Nauk, 02-668 Warszawa, Poland}
\author{Grzegorz Karwasz}
\affiliation{Instytut Fizyki, Uniwersytet Miko{\l}aja Kopernika, 87-100 Toru\'n, Poland }

\begin{abstract}
We develop a simple model of shape resonances in electron-molecule collisions that is based on the modified effective-range expansion and analytical solutions of the Schr\"odinger equation for the long-range part of the interaction potential. We apply our model to electron scattering on N$_2$ and CO$_2$. The parameters of the effective-range expansion (i.e. the scattering length and the effective range) are determined from experimental, integral elastic cross sections in the 0.1 - 1.0 eV energy range. For both molecular targets our treatment predicts shape resonances that appear slightly higher than experimentally known resonances in total cross sections. Agreement with the experiment can be improved by assuming the position of the resonance in a given partial wave. Influence of quadrupole potential on resonances is also discussed: it can be disregarded for N$_2$ but gets significant for CO$_2$. In conclusion, our model developed within the effective range formalism reproduces well both the very low-energy behavior of the integral cross section as well as the presence of resonances in the few eV range.
\end{abstract}
\maketitle

\section{Introduction}

$^2\Pi$ temporary negative ion states in N$_2$ and CO$_2$ at 2.1 eV and 3.8 eV \cite{Schulz,Boness} respectively, are the best known examples of so-called shape resonances, in which the incoming electron is captured within the potential well of the target, rather than to a specific electronic orbital. This distinction came clearly from the first discovery of the shape (in N$_2$) and Feshbach (in He \cite{Schulz2}) resonances by Schulz, see detailed discussion in Ref.  \cite{Schulz3,Buckman1}.

The widths and amplitudes of these resonances, as appearing in total cross sections, are well established, see for ex. Ref. \cite{Szmytkowski95,Sun,Kennerly} for N$_2$ and \cite{Szmytkowski,Kwan} for CO$_2$. In both resonances the vibrational excitation contributes significantly to the total cross section (about 1/6 in N$_2$ and 1/3 in CO$_2$, see \cite{Karwasz01}); both targets are non-polar, with negligible direct rotational excitation cross sections.

However, apart from the resonances, the low-energy dependences of total cross section in N$_2$ and CO$_2$ differ completely: the cross section rises sharply in the zero-energy limit in CO$_2$ \cite{Ferch} and falls slowly in N$_2$ \cite{Sun}. Particularly, the CO$_2$ scattering received a rich theoretical treatment, starting from the early assignment of this state as "shape $^2\Pi_u$ state" \cite{Claydon} and calculations using model potentials, for ex.  by Morrison and co-workers \cite{Morrison77,Collins,Morrison82}. More recent works, both experimental \cite{Allan} and theoretical \cite{Kazansky}, indicate  that the mechanism of this resonance is complex, with coupling between different vibrational modes; the analysis requires the vibrational  excitations to be included \cite{Rescigno02}. However, with few exceptions \cite{Morrison77,Rescigno99} calculations were performed separately for the zero-energy and resonance energy ranges, see for ex. \cite{Morrison82,Rescigno02,Mazevet,Morgan,Domcke,Itikawa}.

In our previous work \cite{Idziaszek} we revisited a rather old problem –- application of the modified effective range theory (MERT) to electron and positron scattering \cite{OMalley}. In that theory the scattering is due to the polarization potential and the short range potential is modelled by an effective range expansion. These features make MERT particularly suitable for the near-to-zero energy range, treated with difficulties in other approaches \cite{Morrison,Saha}. The near-to-zero energy MERT expansion continues to be applied to numerous problems: integral cross sections for atoms \cite{Haddad,Horacek}, non-polar \cite{Willner,Hotop} and polar molecules \cite{Fabrikant83,Vanroose}. MERT is frequently used by experimentalists for extrapolation of measured differential and integral cross sections towards energies and angles inaccessible by experiment \cite{Ferch85,Mann,Karbowski}. Different extensions of MERT, consisting in developing phase shifts (or elements of the reactance, $K$ matrix) into series of $k$ were applied by Morrison et al. \cite{Morrison} to rotational excitation of N$_2$ molecule, and by Macri and Barrachina to electron -– metastable He scattering and electron photodetachment from Li$^-$ \cite{Macri02,Macri03}.

The most extensive, to our knowledge, analysis of MERT applicability in electron scattering was done for noble gases by Buckman and Mitroy \cite{Buckman}. They concluded that MERT fails above about 1 eV, exact energy depending on the target. However, in that as well in numerous other MERT applications, those were integral cross sections which were developed directly in series of the colliding electron momentum $k$. As we showed in the previous work \cite{Idziaszek}, the range of applicability of such a series is limited to a few tenths of eV. Therein \cite{Idziaszek} we proposed a novel way of applying MERT. It consists in calculating the phase shifts due to the polarization potential from Mathieu's solutions of the Schr\"odinger equation and introducing the effective-range expansion {\em exlusively for the short-range potential}. There is no explicit $k$ expansion for the integral cross sections in our approach; only the short-range contribution expressed in terms of an appropriate boundary condition, is expressed as series of $k$. This makes the numerical procedures more complicated but allows to extend the applicability of MERT for positron-–argon and -nitrogen scattering to the range of 2-3 eV. We note that our method closely resembles quantum-defect-theory formulation of the scattering problem, where the quantum-defect parameter, that slowly varies with energy, is expanded in a series of $k$ (see e.g. \cite{Sadeghpour}). In electron-atom scattering the quantum-defect theory has been already applied by Watanabe and Greene \cite{Watanabe} to analyze photodetachment process of K$^{-}$ and by Fabrikant \cite{Fabrikant} to low-energy electron-metal atoms resonances.

In this paper we apply our MERT-based model to the two molecules N$_2$ and CO$_2$, well studied by beam and swarm techniques at energies down to less than 0.1 eV. The parameters of MERT expansion for the short-range potential are derived from fitting procedure using the recommended \cite{Karwasz3} integral cross sections below 1 eV. Next, we use these potentials to calculate cross sections up to 10 eV. In both molecules broad $p$-wave resonances appear. In such an unconstrained fit the resonances appear at higher energies than experimental values. By small, within the experimental error bar, modifications of the data used for the inversion procedure, one can reproduce the exact position of the resonances. In that case both resonances appearing in the $p$ or $d$-partial wave channels are narrower than experimentally determined (but this can be partially caused by neglecting the nuclear motion, see for ex. Ref. \cite{Rescigno02}). We discuss also the relative contributions to integral cross sections from the non-spherical part of the polarizability and the quadrupole moment of molecule, using the distorted-wave approximation.

The paper is organized as follows. In section~\ref{Sec:MERT} we discuss the analytical solutions of the Schr\"odinger equation with the polarization potentials, and we present principles of the modified effective-range expansion. In section~\ref{Sec:Reso} we demonstrate that MERT accounts for shape resonances in $p$ and $d$ waves, assuming some simplified model of electron-molecule potential. The actual electron resonances in N$_2$ and CO$_2$ are analyzed in section~\ref{Sec:Electr}. Section~\ref{Sec:NonIso} discusses correction to MERT cross sections due to the non-spherical part of the polarization potential and the molecule quadrupole moment. We present some conclusions in section~\ref{Sec:Conclu}. Finally, three appendices give some technical details of the analytical solutions for the polarization potential.

\section{Scattering of a charged particle on a molecule and the quantum-defect approach}
\label{Sec:MERT}

The relative motion of a charged particle and a non-polar molecule is described by the Schr\"odinger equation
\begin{equation}
\label{Schr}
\left[- \frac{\hbar^2}{2 \mu} \Delta + V(\mbf{r}) -E  \right]\Psi(\mbf{r}) = 0,
\end{equation}
where $\mu$ is the reduced mass, $E$ is the relative energy of the particles, and $V(\mbf{r})$ is the particle-molecule potential. $V(\mbf{r})$ can be written as a sum
\begin{equation}
V(\mbf{r})= V_\mrm{1}(\mbf{r}) + V_\mrm{2}(\mbf{r}) + V_\mrm{S}(r)
\end{equation}
of long-range contributions
\begin{align}
V_\mrm{1}(\mbf{r}) & = - \frac{\alpha e^2}{2 r^4} \\
V_\mrm{2}(\mbf{r}) & = - \left(\frac{\alpha_2 e^2}{2 r^4} + \frac{Q}{r^3}\right) P_2(\cos \theta)
\end{align}
and a short-range potential $V_{S}(r)$. The long-range potential is expressed in terms of the static spherical and non-spherical polarizabilities $\alpha$ and $\alpha_2$, respectively, and the quadrupole moment $Q$. The short-range part $V_\mrm{S}(r)$ comes into play at the distances comparable to the size of the molecule, where the molecule cannot be treated as a single object. Numerous approaches were proposed in the past, based on modifications of the polarization potential~\cite{Czuchaj}, deriving the short-range interaction from electronic densities etc.~\cite{Gianturco}, or expanding the scattering amplitude at low energies \cite{Fabrikant}.
In the present approach we express $V_\mrm{S}(r)$ in terms of some boundary conditions imposed on the wave function at $r \rightarrow 0$, while the effects resulting from the finite range of  $V_\mrm{s}(r)$ are included explicitly in the framework of the modified effective-range theory.

In the following we assume that the nonisotropic part $V_\mrm{2}(\mbf{r})$ can be neglected in comparison to the isotropic polarization potential $V_\mrm{1}(\mbf{r})$. This is justified as long as the quadrupole moment $Q$ and the nonspherical polarizability $\alpha_2$ expressed in atomic units are much smaller than $\alpha_0$. In such a case the main effect of the non-isotropic terms is the coupling between different partial waves, while their contribution to the total cross-section at low energies remains small. As we show later in section Sec.~\ref{Sec:NonIso}, where we analyze the corrections arising from $V_\mrm{2}(\mbf{r})$, this is a good approximation for N$_2$ and CO$_2$ molecules. The MERT analysis for these molecules is performed in Sec.~\ref{Sec:Electr}.

The radial part of the Schr\"odinger equation with the isotropic polarization potential $V_1(\mbf{r})$ reads
\begin{equation}
\label{RadSchr}
\left[\frac{\partial^2}{\partial r^2} - \frac{l(l+1)}{r^2}
+ \frac{(R^\ast)^2}{r^4} + \frac{2 \mu E}{\hbar^2} \right]\Psi_l(r) = 0,
\end{equation}
where $\Psi_l(r)$ denotes the radial wave function for the partial wave $l$. For the polarization potential, it is convenient to introduce some characteristic units $R^\ast$ and $E^\ast$, where $R^\ast \equiv \sqrt{ \alpha e^2 \mu /\hbar^2}$ denotes the characteristic length of the $r^{-4}$ potential, and $E^{\ast} = \hbar^2/(2 \mu {R^{\ast}}^2)$  is the characteristic energy. With appropriate change of variables, the Schr\"odinger equation \eqref{RadSchr} can be transformed into Mathieu's differential equation of the imaginary argument \cite{Vogt,OMalley,Spector}, and solved analytically in terms of the continued fractions. Some basic properties of the analytical solutions are discussed in Appendices~\ref{App:Sol}, \ref{App:Asympt}, and \ref{App:Exp} (see \cite{Erdelyi,Abramowitz} for more details on Mathieu functions). Here, we focus only on the behavior at small and large distances. For $r \ll R^{\ast}$, when the polarization potential dominates over centrifugal potential and the constant energy term, behavior of $\Psi_l(r)$ is given by
\begin{equation}
\label{Psi1}
\Psi_l(r)  \stackrel{r \rightarrow 0}{\sim} r \sin\left(\frac{R^\ast}{r} + \phi_l\right),
\end{equation}
where $\phi_l$ is a short-range phase, which is determined by $V_\mrm{s}(r)$.
For $E=0$ and $l=0$ the solution \eqref{Psi1} becomes exact at all distances, and from its asymptotic behavior at large distances one can easily determine the value of the $s$-wave scattering length
\begin{equation}
a_s =  - R^{\ast} \cot(\phi_0).
\end{equation}
At large distances: $r \gg R^{\ast}$, $\Psi_l(r)$ must take the form of the scattered wave
\begin{equation}
\label{Psi2}
\Psi_l(r) \stackrel{r \rightarrow \infty}{\sim} \sin( {\tst k r - l \frac \pi 2 + \eta_l} ),
\end{equation}
where $k=\sqrt{2 \mu E}/\hbar$.
Using the Mathieu functions one can find the following relation between the phase shift $\eta_l$ and the short range phase $\phi_l$ \cite{OMalley}
\begin{equation}
\label{taneta}
\tan \eta_l = \frac{m^2 - \tan \delta^2 + \tan (\tst \phi_l + l \frac \pi 2) \tan \delta (m^2 - 1)}
{\tan \delta (1 - m^2) +  \tan (\tst \phi_l + l \frac \pi 2) (1- m^2 \tan^2 \delta)},
\end{equation}
where $\delta = \frac \pi 2 (\nu -l -\frac 1 2)$, and $m$ and $\nu$ are some parameters, that are determined by the analytical solutions (see Appendices~\ref{App:Sol} and \ref{App:Asympt} for details).

In general, the parameter $\phi_l$ entering the asymptotic formula \eqref{Psi1}, depends on energy, and can be expanded in powers of $k$. In our case it is more convenient to expand $\tan(\phi_l+l \frac \pi 2)$, entering formula \eqref{taneta}:
\begin{equation}
\tan(\phi_l+l {\tst \frac \pi 2}) = A_l + \frac12 R^\ast R_l k^2 + \ldots,
\label{ExpEff}
\end{equation}
where $A_l \equiv \left.\tan(\phi_l+l \frac\pi2)\right|_{q=0}$. The lowest order correction in $k$ is quadratic, and can be interpreted as an effective range $R_l$ for the partial wave $l$ \cite{OMalley}.

Combining \eqref{taneta}, \eqref{ExpEff}, \eqref{Expnu}, \eqref{Expm} one can easily obtain the low-energy expansion of the phase shifts \cite{OMalley}
\begin{align}
q \cot \eta_0 & = - \frac{1}{a} + \frac{\pi}{3 a^2} q + \frac{4}{3 a} \ln \left(\frac q 4 \right) q^2
+ \frac{{R_0}^2}{2} q^2 \nonumber \\
& + \left[ \frac \pi 3 + \frac{20}{9a} - \frac{\pi}{3 a^2} - \frac{\pi^2}{9 a^3}
- \frac{8}{3a} \psi({\tst \frac 32}) \right] q^2 + \ldots
\label{ExpEta0} \\
\tan \eta_l & = \frac{\pi q^2}{8(l-\frac12)(l+\frac12)(l+\frac32)}+\ldots , \quad l \geq 1
\label{ExpEta2}
\end{align}
Here, $q = k R^{\ast}$, $a = a_{s}/R^{\ast} = - 1/A_0$, and $b = A_1$. The low-energy expansions \eqref{ExpEta0}-\eqref{ExpEta2} are applicable only at small energies $E \ll e^{\ast}$. In contrast, in our approach we use the exact formula \eqref{taneta} for the phase shift, with the short-range phase expanded according to \eqref{ExpEff}.

\section{Model calculation}
\label{Sec:Reso}

\begin{figure}
	 \includegraphics[width=6cm,clip]{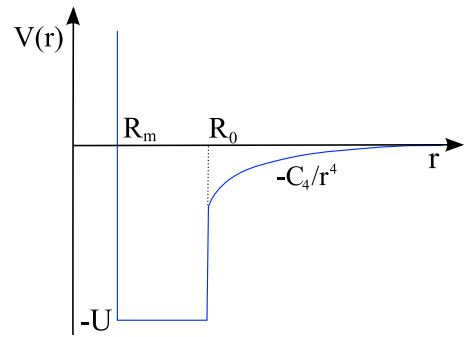}
	 \caption{The model potential Eq.\eqref{Vmod} that is used to test MERT for the shape resonances
     \label{fig:Vmod}}
\end{figure}

To illustrate how MERT describes shape resonances for the long-range $r^{-4}$ potential we consider a simple model of the potential with a square-well interaction at short distances (see Fig.\ref{fig:Vmod})
\begin{equation}
\label{Vmod}
V(r) = \begin{cases}
\infty & r<R_m \\
-U & R_m < r < R_0  \\
- C_4/r^4 & r > R_0
\end{cases}
\end{equation}
Here, $R_m$ denotes a hard-core radius and $R_0$ is the radius of the square-well potential. The potential $V(r)$ does not reproduce the full behavior of the electron-molecule interaction, however, it captures the basic properties of shape resonances in such a system, and allows to verify the accuracy of our MERT-based approach. The Schr\"odinger equation for the potential $V(r)$ and partial-wave quantum number $l$ can be solved exactly:
\begin{equation}
\Psi(r) = \begin{cases}
0,\quad r\squiz{<}R_m \\
 r j_l(\chi r) + C r y_l (\chi r),\quad R_m \squiz{<} r \squiz{<} R_0 \\
 A \sqrt{r} M_\nu\!\!\left(\ln \tfrac{\sqrt{R^\ast}}{\sqrt{k}r}\right) + B \sqrt{r} T_\nu\!\!\left(\ln \tfrac{\sqrt{R^\ast}}{\sqrt{k}r}\right), \quad r\squiz{>}R_0
\end{cases}
\end{equation}
where $\hbar^2 \chi^2/(2 \mu) = E+U$ and constants $A,B,C$ are to be determined from the continuity conditions for $\Psi(r)$ at $r=R_m$ and $r=R_0$. With the help of the asymptotic formulas \eqref{ae3} and \eqref{ae4} for $M_\nu(z)$ and $T_\nu(z)$ for large and negative arguments (large $r$) we find the phase shifts and the total cross-section for scattering on $V(r)$. We compare the exact results with our MERT model where the MERT coefficients $A_l$ and $R_l$ are found by expanding the formula $\tan(\phi_l+\nu\frac{\pi}{2}+\frac{\pi}{4}) = A/B$ (c.f. Eq.~\eqref{comb}) in the powers of $q$, and utilizing series expansion \eqref{Expnu} for the characteristic exponent.

\begin{figure}
	 \includegraphics[width=8.6cm,clip]{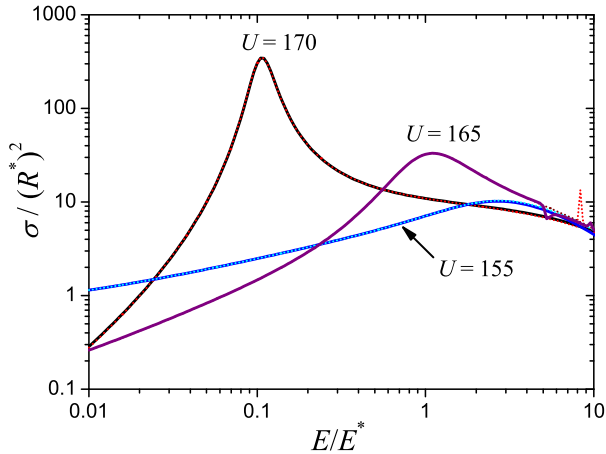}
	 \caption{Total elastic cross-section versus energy for the scattering in the model potential \eqref{Vmod}, calculated for $R_m =0.1 R^{\ast}$, $R_0 =0.5 R^{\ast}$ and for different depths $U$ of the square-well potential. Results are scaled by the the characteristic distance $R^{\ast}$ and the characteristic energy $E^{\ast}$ of the polarization potential. The peaks correspond to the p-wave shape resonances, that are accurately reproduced by MERT.
	 \label{fig:tc}
	 }
\end{figure}

Fig.~\ref{fig:tc} shows the dependence of the total elastic cross-section on energy, for few particular values of $U$, for which  $p$-wave shape-resonances appear. The range $R_0$ was chosen to be smaller than $R^\ast$, but still, of the same order as $R^\ast$, that roughly corresponds to the conditions of the low-energy electron(positron)-molecule scattering. The resonances show up as peaks at some particular values of the energy. We observe that MERT, including only two lowest expansion coefficients $A_l$ and $R_l$ of the short-range potential, reproduces very accurately the exact results. The accuracy decreases for higher energies, where one can observe some discrepancies, in particular small peaks corresponding to the $g$-wave resonances that appear at different energies for exact and MERT curves. This results from higher sensitivity of the quasi-bound states on the parameters of the short-range potential for larger $l$, and the absence of the higher order terms in expansion \eqref{ExpEff} when $E$ becomes too large. Nevertheless, the approximate MERT treatment remains very accurate for the main resonances in $p$-wave channel, even for energies significantly larger than $E^\ast$.

\begin{figure}
	 \includegraphics[width=8.6cm,clip]{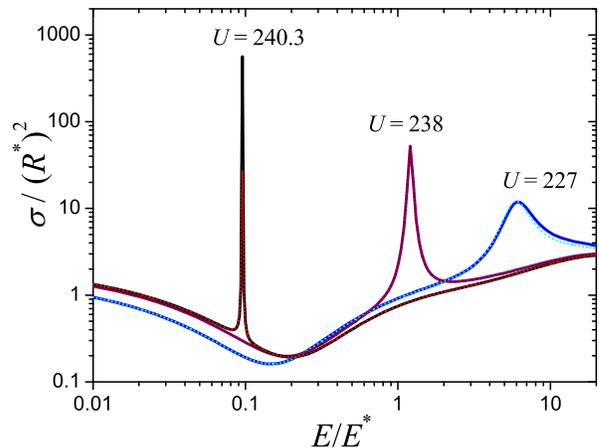}
	 \caption{Same as Fig.~\ref{fig:tc}, but for values of $U$ for which shape resonances in $d$ wave appear.
	 \label{fig:tc2}
	 }
\end{figure}

A similar comparison, but for $d$-wave resonances is presented in Fig.~\ref{fig:tc2}. In this case the resonance peaks are narrower due to the stronger centrifugal barrier, and weaker coupling between quasi-bound and scattering states. Again, the agreement between MERT and the exact results is very good, and some discrepancies can be observed only at $E \sim 10 E^\ast$. Finally, we note that for $l=2$ and parameters of Fig.~\ref{fig:tc2} the quasi-bound state is totally localized within the range of the short-range potential. This shows that MERT is able to reproduce shape resonances resulting solely from the short-range potential, even if the latter is treated in terms of an effective-range expansion.

\section{MERT analysis of low-energy electron scattering}
\label{Sec:Electr}

We investigate the low-energy  electron scattering on N$_2$ and CO$_2$ molecules using a theoretical model based on MERT for $s$ and $p$ partial waves \cite{Idziaszek}. Our model contains four unknown parameters: the scattering length $a=a_s/R^\ast=-1/A_0$ and the effective range $R_0$ for $s$ wave, and the zero-energy contribution $A_1$ and the effective range $R_1$ for $p$ wave, that are determined by fitting to the experimental data. In this way for $l=0$ and $l=1$ we neglect corrections of the order higher than $q^2$ in expansion \eqref{ExpEff}. On the other hand, for partial waves with $l \geq 2$ we include only the lowest-order contribution to the scattering phase shifts that are due to the long-range polarization potential \eqref{ExpEta2}. This is sufficient as long as the energy $E$ remains smaller than the centrifugal barrier, that for $l=2$ has height of $9 E^{\ast}$. The total elastic cross-section $\sigma(k) = 4 \pi k^{-2} \sum_{l} \sin^2 \eta_l(k)$ is determined from the phase shifts \eqref{taneta} and \eqref{ExpEta2}, with $\nu$ and $m$ evaluated numerically, using the procedure described in Appendices~\ref{App:Sol} and \ref{App:Asympt}.

\begin{figure}
	 \includegraphics[width=8.6cm,clip]{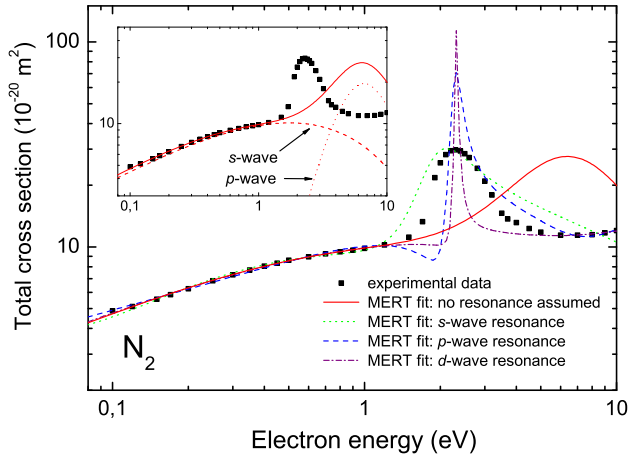}
	 \caption{ Total cross-section for the scattering of electrons on N$_2$ versus the energy. Depicted are: recommended experimental data from review \cite{Karwasz3} (stars), the theoretical fits based on MERT, assuming the resonance at 2.1eV in $s$ wave (dotted line), $p$ wave (dashed line), and $d$ wave (dot-dashed line), and without assumption with respect to the position of the resonance (solid line).
The inset shows in addition the $s$-wave and $p$-wave contributions to the MERT fit not assuming the position of resonance.
	 \label{fig:N2}
	 }
\end{figure}

\begin{figure}
	 \includegraphics[width=8.6cm,clip]{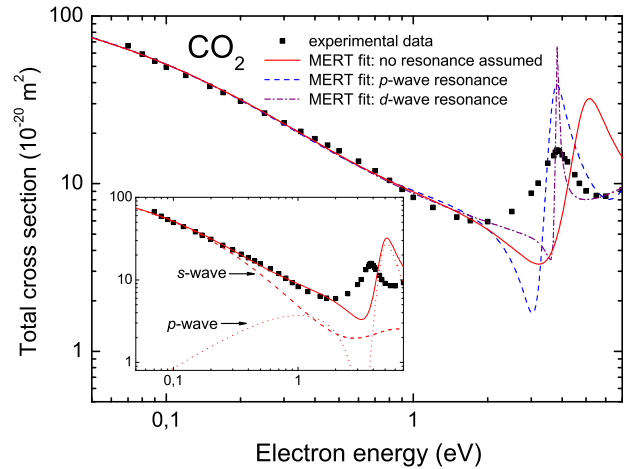}
	 \caption{ The same as fig.~\ref{fig:N2} but for the scattering of electrons on CO$_2$ with the recommended experimental data from review \cite{Karwasz3}. Note, missing curve corresponding to $s$-wave resonance, which does not fit well to the experimental data in this case.
	 \label{fig:CO2}
	 }
\end{figure}

In the calculations we use recent experimental values of the polarizability, as measured in electron scattering experiments: $\alpha=11.54 {a_0}^3$ for N$_2$ and $\alpha=16.92 {a_0}^3$ for CO$_2$ \cite{Olney}. First, we fit our model by the least-square, not weighted method to the experimental data exclusively below the resonances, i.e. for energies $E \leq 1.2$eV and $E \leq 2$eV for N$_2$ and CO$_2$, respectively, down to the lowest energies available from beam experiments \cite{Karwasz3}. The fitted parameters from this check are listed in Table~\ref{Tab1}. Note, that the scattering potential parameters are expressed in characteristic distance units. For example, in N$_2$ the scattering length $a_s$ amounts to $0.404$ atomic units, close to the value of $0.420$~a.u. derived by Morrison et al. \cite{Morrison}. Subsequently, we extend MERT analysis, using the fitted parameters to higher energies. Figs.~\ref{fig:N2} and \ref{fig:CO2} compare the experimental data (points) with the theoretical curves obtained from this fit (solid lines). For both molecules resonance maxima appear in the scattering cross-section, however they are located at somewhat higher energies that the experimental ones (about 6 eV for N$_2$ and 5 eV for CO$_2$). For both molecules these ``spontaneously'' appearing resonances are due to the $p$-wave, see inserts in Fig.~\ref{fig:N2} and \ref{fig:CO2}. In the case of CO$_2$ the $p$-wave is subject to rapid changes also below the resonance. This is to be attributed to large values of $b$ and $R_1$ of the potential (see table \ref{Tab1}). Note that in experiment \cite{Allan1} some vibrational modes (001) show resonant maxima below the main 3.6 eV peak.

\begin{table}
\begin{ruledtabular}
\begin{tabular}{lllllll}
 & $R^{\ast}$($a_0$) & $E^{\ast}$(eV) & $a$ & $A_1$ & $R_0/R^{\ast}$ & $R_1/R^{\ast}$ \\
\hline
N$_2$ & 3.397 & 1.179 & 0.119 & -0.537 & 0.110 & 0.262 \\
CO$_2$ & 4.113 & 0.8041 & -1.606 & -3.537 & -0.325 & 1.208 \\
\end{tabular}
\end{ruledtabular}
\caption{\label{Tab1} Characteristic distance $R^{\ast}$, characteristic energy $E^{\ast}$ and four fitting parameters: $a=a_\mrm{s}/R^\ast$
($s$-wave scattering length), $A_1$ (zero-energy contribution from the short-range potentials for $p$-wave), $R_0$ ($s$-wave effective range), and $R_1$ ($p$-wave effective range) for N$_2$ and CO$_2$ in the case of unconstrained fits and using only the very low energy experimental data.
}
\end{table}

The range of fitted data corresponds to quite low energies in terms of characteristic units: $E\lesssim E^{\ast}$ ($E\lesssim 2.5 E^{\ast}$) for N$_2$ (CO$_2$). Therefore, the fitting procedure, within the error bar of experimental points, is not able to predict with sufficient accuracy the cross section in the energy range corresponding to resonance peaks. Also, neglection of the nonisotropic part of the potential leads to some additional errors. Moreover, our scheme neglects the possibility of $d$-wave or higher partial wave resonances, that can be suggested e.g. by the behavior of differential cross-section for the scattering on N$_2$. To eliminate these shortcomings we have subsequently added to the fitting procedure also some experimental points, located in the energy range above the resonance: $6$eV~$<E< 10$eV (N$_2$), and $5.5$eV~$<E< 7$eV (CO$_2$). In addition we have considered three possible scenario assuming that resonances occur in partial waves $s$, $p$ or $d$, exactly at the energy of the resonance peak observed in the experimental data. This corresponds to the requirement that the phase shift $\eta_l$ assumes value $\pi/2$ at the peak location for $l=0,1,2$, respectively. In this way we obtain some constrains on the six fitting parameters: $a$, $A_1$, $A_2$ (zero-energy contributions) and $R_0$, $R_1$ and $R_2$ (effective ranges). The results are shown in Figs.~\ref{fig:N2} and \ref{fig:CO2}, presenting apart from the previous data, also theoretical curves with resonances assumed in $s$, $p$ and $d$ waves. The fitted parameters are listed in tables~\ref{Tab2} and \ref{Tab3}.

\begin{table}
\begin{ruledtabular}
\begin{tabular}{lllllll}
 &  $a$ & $A_1$ & $A_2$ & $R_0/R^{\ast}$ & $R_1/R^{\ast}$ & $R_2/R^{\ast}$\\
\hline
$s$-wave res. & -0.447 & -0.021 & --- & -3.100 & -0.104 & --- \\
$p$-wave res. & 0.157 & -4.314 & -0.130 & -47.05 & 4.704 & 0.079 \\
$d$-wave res. & 0.123 & -0.390 & -0.222 & -0.301 & -0.174 & 0.293 \\
\end{tabular}
\end{ruledtabular}
\caption{\label{Tab2} Six fitting parameters: $a=a_\mrm{s}/R^\ast$ ($s$-wave scattering length), $A_1$ (zero-energy contribution for $p$-wave), $A_2$ (zero-energy contribution for $d$-wave), $R_0$ ($s$-wave effective range), $R_1$ ($p$-wave effective range) and
$R_2$ ($d$-wave effective range) for N$_2$ in the case of fits assuming shape resonances in $s$, $p$ and $d$ waves. The fits are made to the data below and above the resonance (see text for details). Note that for $s$-wave resonance only four parameters (two partial waves) were sufficient to fit the data.
}
\end{table}
\begin{table}
\begin{ruledtabular}
\begin{tabular}{lllllll}
 &  $a$ & $A_1$ & $A_2$ & $R_0/R^{\ast}$ & $R_1/R^{\ast}$ & $R_2/R^{\ast}$\\
\hline
$p$-wave res. & -1.616 & -6.250 & -0.406 & -0.266 & 2.780 & 0.131 \\
$d$-wave res. & -1.605 & -2.470 & -2.624 & -0.340 & -0.234 & 1.180 \\
\end{tabular}
\end{ruledtabular}
\caption{\label{Tab3} Six fitting parameters: $a=a_\mrm{s}/R^\ast$ ($s$-wave scattering length), $A_1$ (zero-energy contribution for $p$-wave), $A_2$ (zero-energy contribution for $d$-wave),
$R_0$ ($s$-wave effective range), $R_1$ ($p$-wave effective range) and
$R_2$ ($d$-wave effective range) for CO$_2$ in the case of fits assuming shape resonances in $p$ and $d$ waves. The fits are made to the data below and above the resonance (see text for details).
}
\end{table}

We observe that for N$_2$ the height of the resonance peak for $s$ wave agrees well with the experimental data (we show experimental values averaged over the vibrational structure seen in the resonance, following recommended data from ref. \cite{Karwasz3}). This would suggest a possible $s$-wave origin of the resonance. But this point requires some more detailed discussion.

As said before, the generally accepted classification of resonances describes ``shape'' resonances as occurring within the shape of the effective potential, formed from the static and polarization contributions with the centrifugal barrier added. Obviously, for the $s$-wave the centrifugal barrier does not exist so trapping of the electron within the shape of the effective potential is impossible. However, our calculations show a quick change in the s-wave phase shift, occurring approximately in the region of the experimentally observed resonance so following the interpretation by Fano \cite{Fano,Bianconi} we would classify this state as a resonance. On the other hand, assumption of resonances in $p$ and $d$ waves would lead to much narrower and higher peaks in the total cross section than the experimental values. Therefore, a simple interpretation of the the N$_2$ resonance is not easy: i) elastic differential cross sections at the resonance maximum (2.47 eV) show a dominating $p$-wave contribution, ii) differential cross sections for the first vibrational excitation ($\nu$=0-1) at 2.45 eV show almost a perfect $d$-wave dependence \cite{Allan1} while iii) the MERT analysis does not exclude the resonance in the $s$-partial wave. Clearly, several partial waves contribute to the resonance and MERT analysis is not able to separate these contributions.

For CO$_2$ the MERT fit with constraints does not improve much the agreement with the experimental total cross sections, see Fig. 3. The MERT analysis excludes completely the $s$-wave resonance, and both $p$ and $d$ wave resonances differ much in amplitude and width from the experimental values. For CO$_2$ resonance the interplay of different partial waves is even more complex than for N$_2$. On the other hand, recent calculations \cite{Kazansky,Rescigno99} and experiments \cite{Allan} showed that the broad peak in CO$_2$ cross section is rather a superposition of two resonances for different molecular geometries, resulting from the bent configuration in the dominating vibrationally excited mode (010).

\section{Corrections due to the nonisotropic part of the long-range potential}
\label{Sec:NonIso}

Already early works stressed importance of other, apart from the sperical polarization, components of the long-range potential. For both molecules the non-sperical part of the polarizability is rather strong (see table IV); for CO$_2$ also the quadrupole moment is significant. 
In the present analysis we approximate the quadrupole and non-spherical polarizability contributions together, but using two distinct methods for the zero-energy limit and at low $E \sim E^{\ast}$ energies. To analyze the contribution from the nonisotropic part of the potential at very low energies ($E \ll E^{\ast}$), we can compare the small-$k$ series expansions for the whole potential $V_1(r)+V_2(r)$ \cite{Fabrikant84}, to the expansion including the spherical part $V_1(r)$ only. For the isotropic part, the total cross-section for scattering on N$_2$ and CO$_2$ molecules can be obtained from expansions of the phase shifts given by Eqs.~\eqref{ExpEta0}-\eqref{ExpEta2}. Substituting the parameters of the potential listed in the Table~\ref{Tab:Par} one obtains
\begin{align}
\label{sigmaN2}
\frac{\sigma_{\mrm{N}_2}(q)}{4 \pi} & = a^2 +\text{24.60} q a + \text{31.32}q^2 \log (q) a^2 + O(q^3) \\
\label{sigmaCO2}
\frac{\sigma_{\mrm{CO}_2}(q)}{4 \pi} & = a^2 +\text{35.44} q a +\text{45.12} q^2 \log (q) a^2 + O(q^3) \\
\end{align}
On the other hand, application of the small-$k$ expansion for the full potential \cite{Fabrikant84} leads to
\begin{align}
\label{sigmaN2F}
\frac{\sigma_{\mrm{N}_2}(q)}{4 \pi} & = a^2 + \text{24.93} q a+ \text{0.23} q + \text{0.11} \nonumber \\
& \quad + \text{31.63} (a-\text{0.15})(a+\text{0.16}) q^2 \log (q) + O(q^3) \\
\label{sigmaCO2F}
\frac{\sigma_{\mrm{CO}_2}(q)}{4 \pi} & =  a^2+\text{39.56} q a-\text{3.97} q +\text{1.32} \nonumber \\
& \quad + \text{49.09} (a-\text{0.53}) (a+\text{0.56}) q^2 \log(q)+ O(q^3)
\end{align}
We observe that for the typical scales of the scattering length $a \sim 1$, the differences in the coefficients of Eqs.\eqref{sigmaN2F}-\eqref{sigmaCO2F} and Eqs.\eqref{sigmaN2}-\eqref{sigmaCO2} are at most of the order of 10\%.

\begin{table}
\begin{ruledtabular}
\begin{tabular}{llll}
 & $\alpha$($a_0^3$) & $\alpha_2$($a_0^3$) & $Q$($e a_0^2)$ \\
\hline
N$_2$ & 11.54 & 3.08 & -1.09 \\
CO$_2$ & 16.92 & 9.20 & -3.86 \\
\end{tabular}
\end{ruledtabular}
\caption{\label{Tab:Par} Parameters of the long-range potential $V_\mrm{L}(\mbf{r})$ for molecules considered in our analysis. For $\alpha_0$ we use recent experimental values of \cite{Olney}, and for $\alpha_2$ and $Q$ we use experimental values used in Refs. \cite{Morrison} for N$_2$ and in Ref. \cite{Morrison77} for CO$_2$}.
\end{table}

For higher energies $E \sim E^{\ast}$ the importance of the nonisotropic part can be assessed with the help of the distorted wave approximation \cite{Mott}. The scattering amplitude can be written as a sum of the contributions from $V_1$ and $V_2$ separately
(see e.g. \cite{Dewangan})
\begin{equation}
\label{famp}
f(\mbf{k}_\mrm{f},\mbf{k}_\mrm{i}) = - \frac{m}{2 \pi \hbar^2} \left(\langle \mbf{k}_\mrm{f}|V_1|\Phi^{+}(\mbf{k}_\mrm{i})\rangle
+ \langle \Phi^{-}(\mbf{k}_\mrm{f})|V_2|\Psi^{+}\rangle\right),
\end{equation}
where $\Phi^{\pm}(\mbf{k})$ is the scattering state for the isotropic part $V_1$
\begin{equation}
\Phi^{\pm}(\mbf{k}) = e^{i \mbf{k} \mbf{r}} + G_0^{\pm} V_1 \Phi^{\pm}(\mbf{k}),
\end{equation}
$G_0^{\pm}$ is the free propagator
\begin{equation}
G_0^{\pm} = \lim_{\epsilon \rightarrow 0} (E - \mbf{p}^2/2m \pm i \epsilon)^{-1},
\end{equation}
and $\Psi^{+}$ is the scattering state for the full potential, satisfying the the following Lippmann-Schwinger equation
\begin{align}
\Psi^{+} & = \Phi^{+}(\mbf{k}_\mrm{i}) + G_1^{+} V_2 \Psi^{+},\\
G_1^{+} & = \lim_{\epsilon \rightarrow 0} (E - \mbf{p}^2/2m - V_1 + i \epsilon)^{-1}.
\end{align}
The second term in Eq.~\eqref{famp} can be written in the form of a Born series
\begin{equation}
\langle \Phi^{-}(\mbf{k}_\mrm{f})|V_2|\Psi^{+}\rangle = \langle \Phi^{-}(\mbf{k}_\mrm{f})|V_2+V_2 G_1^+ V_2 + \ldots
|\Phi^{+}(\mbf{k}_\mrm{i})\rangle.
\end{equation}
Finally, the total cross-section can be calculated applying the optical theorem
$\sigma =\frac{4 \pi}{k} \mrm{Im} f(\mbf{k}_\mrm{i},\mbf{k}_\mrm{i})$.

After averaging over different orientations of molecules, the lowest order contribution from the nonisotropic part of the potential
$\langle \Phi^{-}(\mbf{k}_\mrm{f})|V_2|\Phi^{+}(\mbf{k}_\mrm{i})\rangle$ vanishes, and for small anisotropic part, the leading contribution is provided by $\langle \Phi^{-}(\mbf{k}_\mrm{f})|V_2 G_1^+ V_2 |\Phi^{+}(\mbf{k}_\mrm{i})\rangle$. We evaluate this term using analytical solutions for the spherical potential $V_1$ and introducing an additional cut-off parameter $R_0$, that describes the distance when the interaction potential takes a form of the long range potential $V_1 + V_2$, which is roughly of the order of the size of the molecule.

Tables \ref{Tab:Corr1} and \ref{Tab:Corr2} list the the corrections due to the nonisotropic part $V_2$, together with
total cross sections coming from the isotropic part $V_1$. The analytical unperturbed solution for $V_1$ was calculated for parameters of Table~\eqref{Tab1}, corresponding to the low-energy fits. For $N_2$ the corrections are small, of the order of 1\%. On the other hand, for CO$_2$, the corrections are as large as 40\% of the total cross-section $\sigma_0$, therefore in this case an accurate analysis would require full inclusion of $V_2$. We note that corrections slightly depend on the cut-off parameter
$R_0$, that is a consequence of the fact that the analytical solutions for the purely polarization potential $V_1$ are not physical at small distances, where the short-range part of the potential comes into play. Therefore corrections listed in tables \ref{Tab:Corr1} and \ref{Tab:Corr2} estimate contributions of $V_2$ from large distances only, where the potential takes the asymptotic form $V_1 + V_2$. 

\begin{table}
\begin{ruledtabular}
\begin{tabular}{lllll}
$E/E^{\ast}$ & $\sigma_0$ & $R_0=0.5 R^{\ast}$ & $R_0 = 0.75 R^{\ast}$ & $R_0 = R^{\ast}$ \\
\hline
1.0 & 3.145 & 0.0306 & 0.0311 & 0.0300 \\
2.0 & 4.014 & 0.0147 & 0.0135 & 0.0138 \\
4.0 & 7.428 & 0.0059 & 0.0052 & 0.0053 \\
\end{tabular}
\end{ruledtabular}
\caption{\label{Tab:Corr1} The total elastic cross-section $\sigma_0$ and corrections due to the nonspherical part of the long-range potential in N$_2$ molecule, for different energies $E$ of the scattered electron and different short-range cut-off parameters $R_0$.}
\end{table}

\begin{table}
\begin{ruledtabular}
\begin{tabular}{lllll}
$E/E^{\ast}$ & $\sigma_0$ & $R_0=0.5 R^{\ast}$ & $R_0 = 0.75 R^{\ast}$ & $R_0 = R^{\ast}$ \\
\hline
1.0 & 2.185 & 0.174 & 0.796 & 0.698 \\
2.0 & 1.345 & 0.455 & 0.514 & 0.363 \\
4.0 & 0.693 & 0.207 & 0.213 & 0.153 \\
\end{tabular}
\end{ruledtabular}
\caption{\label{Tab:Corr2} The total elastic cross-section $\sigma_0$ and corrections due to the nonspherical part of the long-range potential in CO$_2$ molecule, for different energies $E$ of the scattered electron and different short-range cut-off parameters $R_0$.}
\end{table}

\section{Conclusions}
\label{Sec:Conclu}

MERT analytical solution was applied previously \cite{Idziaszek} for positron scattering cross sections. Those cross sections fall monotonically with energy, up to the positronium formation threshold, and do not exhibit particular structures, see \cite{Karwasz2}. The model approximated well Ar and N$_2$ integral elastic cross sections at 0-2~eV. The derived scattering potentials were characterized by negative values of the scattering length-like parameters: $a$ and $A_1$, for $s$ and $p$ waves, respectively.

At present, the MERT analysis was applied for electron scattering on non-polar molecules, N$_2$ and CO$_2$. Parameters of the short range potential (the scattering length and the effective range) for the $s$, $p$, and for higher energies also the $d$ partial wave, were fitted. The model approximates very well the experimental data up to about 1-2~eV. The $s$-wave scattering length deduced is positive for N$_2$ and negative for CO$_2$. Deriving the potential parameters from the experimental data below resonance and using them again for our model at higher energies, we obtained $p$-wave resonances for both molecules, although at slightly higher energies than observed experimentally. To check for the consistency of the fit, the model was applied jointly to the data below and above resonance, assuming that the resonance occur the in $s$, $p$ or $d$ wave, at the energy as observed experimentally. Within the latter constraint, for N$_2$ a ``resonance-like'' peak, corresponding to the $\pi/2$ phase change appears also in the $s$-wave channel. In that case the peak is not as sharp as pure shape resonances in $p$ and $d$ partial waves.

As far as amplitude are considered, in N$_2$ fixing the resonance position reproduces well the absolute value of the total cross section in its maximum, but only if the resonance is assumed to appear in the $s$-wave channel. Such a fixed fit does not improve the agreement in the case of CO$_2$. More experimental points at low energies, also for N$_2$, would be useful for such a modelling. By the way, we recall that Ramanan and Freeman \cite{Ramanan}, on the basis of their swarm experiment even stated: ``experimental data presently available do not rule out the existence of a Ramsauer-Townsend minimum at an electron energy of 0.4~meV or lower''.

Summarizing, the present work unifies in a single model the very low-energy dependence of the integral cross sections in N$_2$ and CO$_2$ with the occurrence of shape resonances at a few eV energies. It confirms that limitations on applicability of MERT to the very low energy range, come rather from direct expansions of the phase-shifts in series of $k$, rather than from any intrinsic defects of the effective range model. Obviously, several aspects of resonant scattering, like partitioning between the elastic and vibrational/rotational excitation are beyond the capacities of our model. It would be challenging to incorporate MERT potentials to more advanced theoretical treatments, in particular of inelastic direct and resonant scattering \cite{Mazevet,Rescigno02,Rescigno99}.

\acknowledgments

The authors thank L.P. Pitaevskii for stimulating discussions. One of the author (ZI) acknowledges support of the Polish Government Research Grant for 2007-2010.

\appendix

\section{Solutions of the Schr\"odinger equation for $1/r^4$ potential}
\label{App:Sol}
The method for solving the Schr\"odinger equation with the polarization potential was discussed in our previous paper \cite{Idziaszek}. Below we indicate the basic steps of the derivation. In the radial Schr\"odinger equation \eqref{RadSchr} we substitute $r= \sqrt{R^\ast} e^{-z} / \sqrt{k}$ and $\Psi_l(r) = \psi(r) \sqrt{r/R^{\ast}}$, which yields the Mathieu's modified differential equation
\cite{Erdelyi,Abramowitz}
\begin{equation}
\label{Mathieu}
\frac{d^2 \psi}{d z^2} - \left[ a - 2 q \cosh 2 z \right] \psi = 0.
\end{equation}
where $a=(l+{\tst \frac 12})^2$ and $q=k R^{\ast}$. Two
linearly independent solutions $M(z)$ and $T(z)$ can be expressed in the following form \cite{Erdelyi,Spector}
\begin{eqnarray}
\label{DefM}
M_\nu(z) & = & \sum_{n=-\infty}^{\infty} (-1)^n c_n(\nu) J_{2n + \nu} (2 \sqrt{q} \cosh z), \\
\label{DefT}
T_\nu(z) & = & \sum_{n=-\infty}^{\infty} (-1)^n c_n(\nu) Y_{2n + \nu} (2 \sqrt{q} \cosh z),
\end{eqnarray}
Here, $\nu$ denotes the characteristic exponent, and $J_{\nu}(z)$ and $Y_{\nu}(z)$ are Bessel and Neumann functions respectively. Substituting the ansatz \eqref{DefM} and \eqref{DefT} into \eqref{Mathieu} one obtains the following recurrence relation:
\begin{equation}
\label{rec}
\left[(2n+\nu)^2 - a \right] c_n + q (c_{n-1} + c_{n+1}) = 0,
\end{equation}
which can be solved in terms of continued fractions. To this end we introduce $h_n^{+} = c_n / c_{n-1}$ and $h_n^{-} = c_{-n} / c_{-n+1}$ for $n > 0$, which substituted into \eqref{rec} gives the continued fractions
\begin{equation}
\label{rec1}
h^{+}_{n} = - \frac{q}{q h^{+}_{n+1} + d_n}, \qquad h^{-}_{n} = - \frac{q}{q h^{-}_{n+1} + d_{-n}},
\end{equation}
with $d_n = (2n+\nu)^2 - a$. Characteristic exponent has to be determined from Eq.~\eqref{rec} for $n=0$ expressed in terms of
$h^{-}_{1}$ and  $h^{+}_{1}$. In practice, to find numerical values of the coefficients $c_n$ we set $h^{+}_{m}=0$  and $h^{-}_{m}=0$ for some, sufficiently large $m$ and calculate $h^{+}_{n}$ and $h^{-}_{n}$ up to $n=1$.
using \eqref{rec1}.

\section{Asymptotic expansions for large arguments}
\label{App:Asympt}
Asymptotic behavior of $M_\nu(z)$ and $T_\nu(z)$ for large $z$ follows immediately from asymptotic expansions of Bessel functions for large arguments \cite{Abramowitz}
\begin{align}
\label{ae1}
M_\nu(z) \stackrel{z \rightarrow \infty}{\longrightarrow} & \sqrt{\frac 2 \pi} \frac{e^{-z/2}}{q^{1/4}} s_\nu
\cos \left( {\tst e^{z} \sqrt q - \frac \pi 2 \nu  - \frac \pi 4 }\right) \\
\label{ae2}
T_\nu(z) \stackrel{z \rightarrow \infty}{\longrightarrow} & \sqrt{\frac 2 \pi} \frac{e^{-z/2}}{q^{1/4}} s_\nu
\sin \left( {\tst e^{z} \sqrt q - \frac \pi 2 \nu  - \frac \pi 4 }\right)
\end{align}
where $s_\nu = \sum_{n=-\infty}^{\infty} c_n (\nu)$. To obtain asymptotic behavior for large and negative $z$ it is necessary to connect solutions $M_\nu(z)$ and $T_\nu(z)$, with another pair of solution $M_{\nu}(-z)$ and $T_{\nu}(-z)$ across $z=0$ \cite{Spector}. In this way one obtains \cite{Idziaszek}
\begin{align}
\label{ae3}
M_\nu(z) \stackrel{z \rightarrow - \infty}{\longrightarrow} & \sqrt{\frac 2 \pi} \frac{e^{z/2}}{q^{1/4}} m s_\nu
\cos \left( {\tst \sqrt q  e^{-z} + \frac \pi 2 \nu  - \frac \pi 4 }\right) \\
T_\nu(z) \stackrel{z \rightarrow - \infty}{\longrightarrow} & - \sqrt{\frac 2 \pi} \frac{e^{z/2}}{q^{1/4}}
\frac{s_\nu}{m}
\Big[ \sin \left( {\tst \sqrt q e^{-z} + \frac \pi 2 \nu  - \frac \pi 4 }\right) \nonumber \\
\label{ae4}
& - \cot \pi \nu (m^2-1) \cos \left( {\tst \sqrt q e^{-z} + \frac \pi 2 \nu  - \frac \pi 4 }\right) \Big]
\end{align}
where $m = \lim_{z \rightarrow 0^{+}} M_\nu(z) / M_{-\nu}(z)$. To evaluate $m$ numerically, we rather avoid using formula \eqref{DefM}, where summation over $n$ converges very slowly for $z \rightarrow 0$. An alternative approach is to use a different representation for the solutions of the Mathieu's equation \cite{Erdelyi}
\begin{equation}
W_\nu(z) = \sum_{n=-\infty}^{\infty} c_n(\nu) e^{2n +\nu},
\end{equation}
where $c_n(\nu)$ are the same coefficients as used in definition of $M_\nu(z)$ and $T_\nu(z)$. One can show that $W_\nu(z)$ differs from $M_\nu(z)$ only by some constant prefactor. We utilize this fact calculating the prefactor at some moderate values of the argument ($z \sim 1$), and then evaluate $m$ using the function $W_\nu(z)$, that has well defined behavior at $z \rightarrow 0$.

Finally the radial wave function $\Psi_l(r)$ for partial wave $l$ can be written as linear combination of $M_\nu(z)$ and $T_\nu(z)$
\begin{align}
\label{comb}
\Psi_l(r) = & \sin ({\tst \phi_l + \frac \pi 2 \nu  + \frac \pi 4}) \sqrt{\frac{R^{\ast}}{r}}
M_\nu\left( \ln \frac{ \sqrt{R^{\ast}}}{\sqrt{k} r} \right) \nonumber  \\
& + \cos ({\tst \phi_l + \frac \pi 2 \nu  + \frac \pi 4}) \sqrt{\frac{R^{\ast}}{r}}
T_\nu\left( \ln \frac{ \sqrt{R^{\ast}}}{\sqrt{k} r} \right),
\end{align}
where $\phi_l$ is a parameter which appear in the small $r$ expansion \eqref{Psi1}. Now, the behavior of $\Psi(r)$ at small and large distance described by Eqs.~\eqref{Psi1}-\eqref{taneta}, can be readily
obtained from asymptotic expansions \eqref{ae1}-\eqref{ae4}.

\section{Expansions for small energies}
\label{App:Exp}
Behavior of the parameters $m$ and $\nu$ at small energies is given by the following series expansions \cite{Meixner}
\begin{align}
\nu(q) = & l + {\tst\frac{1}{2}}  - \frac{ q^2}{4(l-\frac12)(l+\frac12)(l+\frac32)} +O(q^4),
\label{Expnu}
\end{align}
and
\begin{align}
m(q) = & \left(\frac{q}{4}\right)^{l+\frac12} \frac{\Gamma(\frac12-l)}{\Gamma(\frac32+l)}+ O(q^{l+5/2}),
\label{Expm}
\end{align}
where $\psi(x)$ denotes the digamma function \cite{Abramowitz}. To obtain the above formulas, one can in fact solve the recurrence relation \eqref{rec} keeping only the lowest-order terms $c_0$, $c_{-1}$ and $c_1$.

\end{document}